\documentclass{amsart}
\usepackage{amssymb}
\usepackage[pdftex]{graphicx}
\usepackage{subfigure}
\usepackage{mathrsfs}
\usepackage{setspace}

\DeclareGraphicsExtensions{eps}

\begin{document}

\title{A Fast $\mathcal{L}_p$ Spike Alignment Metric}
\author{Alexander J. Dubbs $^{12}$, Brad A. Seiler $^{12}$, and Marcelo O. Magnasco$^1$ }

\maketitle

{$^1$Center for Studies in Physics and Biology, Rockefeller University, New York NY

$^2$Harvard Faculty of Arts and Sciences, Cambridge MA}

\begin{abstract}
The metrization of the space of neural responses is an ongoing research program seeking to find natural ways to describe, in geometrical terms, the sets of possible activities in the brain. One component of this program are the {\em spike metrics}, notions of distance between two spike trains recorded from a neuron.  Alignment spike metrics work by identifying ``equivalent'' spikes in one train and the other.
We present an alignment spike metric having $\mathcal{L}_p$ underlying geometrical structure; the $\mathcal{L}_2$ version is Euclidean and is suitable for further embedding in Euclidean spaces by Multidimensional Scaling methods or related procedures. We show how to implement a fast algorithm for the computation of this metric based on bipartite graph matching theory. 
\end{abstract}

\section{Introduction}

The analysis of neural signals seeks to ``translate'' any set of neural impulses into a language we understand \cite{riekeetal1997}.  But as the neural signals elicited by the same stimulus are never exactly alike, we need a quantitative means of determining when two neural signals serve the same purpose. In what follows, we shall restrict our attention to signals from isolated cell electrophysiology, and represent a ``spike train'' neural signal as a marked point process, an ordered list of spike times labeled by their respective neurons of origin \cite{aronovvictor2004}. To compare spike trains, Victor and Purpura introduced the notion of a \textit{spike metric}, a distance function on the set of spike trains endowing it with the topological properties of a metric space \cite{victorpurpura1996}, \cite{victorpurpura1997}. Spike metrics have been successfully used to quantify variability in data and characterize neural coding in the visual, auditory, olfactory, taste, and electric sensory systems \cite{victor2005}. Numerous spike metrics have been proposed with the goals of clustering neural signals by stimulus \cite{victorpurpura1997}, \cite{schrauwencampenhout2007} and embedding them in Euclidean space \cite{victorpurpura1997} via multidimensional scaling (MDS) \cite{kruskalwish1978}. MDS allows one to visualize the geometry of spike data, especially after the use of sophisticated dimensionality-reduction procedures such as Local Linear Embedding \cite{roweissaul2000}. It also permits us to use more sophisticated clustering techniques designed to work only in vector spaces, such as PCA-guided K-means \cite{dinghe2004} and soft-margin support vector machines, as done (without embedding) in \cite{schrauwencampenhout2007}.

One approach to spike metric design for spike trains generated by single neurons uses the rate-coding hypothesis of spike generation, the idea that spike trains induced by the same stimulus are instances of the same variable-rate Poisson process \cite{riekeetal1997}. The implication is that to compare two spike trains, one must compare their estimated underlying rate functions. To estimate the rate function of a spike train, represent it as a sum of Kronecker delta functions at the spike times and stream it through a low-pass filter, or, equivalently, convolve it with a nonnegative function, or \textit{kernel}, that integrates to $1$. Common choices of kernels include the one-sided exponential tail \cite{vanrossum2001}, the two sided exponential tail, the `tent' function ($0$ above and below a certain range, and linearly increasing and then decreasing at the same rate within that range), and the normal distribution function \cite{schrauwencampenhout2007}. The value of the metric between two spike trains is the $\mathcal{L}_p$ norm of the difference between their estimated rate functions ($p \geq 1$). For $p = 2$, such metrics resemble Euclidean distances, and MDS can embed them in (mostly) Euclidean finite dimensional vector spaces.

However, their is substantial evidence that the time-coding of neural signals contains meaning beyond that conveyed by the firing rate \cite{chaseyoung2006}, \cite{gerstneretal1997}, \cite{riekeetal1997}, \cite{samondsbonds2003}. Based on this fact, Victor and Purpura introduced a spike metric equal to the minimum `cost' of aligning the spikes in two spike trains, \cite{victorpurpura1996}, \cite{victorpurpura1997}. Their approach was to generalize methods used for comparing strands of DNA. In particular, they reengineered Sellers' dynamic programming algorithm for aligning and calculating the evolutionary distance between pairs of DNA sequences \cite{sellers1974} to compute a distance between pairs of spike trains \cite{victorpurpura1996}, \cite{victorpurpura1997}. This metric preserves the integrity of individual spikes instead of viewing them as contributions to a rate function. It has the added advantage of being able to compute distances between spike trains that may have contributions from multiple neurons, and in the future, it may be generalized to align spikes from multiple spike trains at once, as has been done with DNA \cite{notredame2002}. The problem with it is that it resembles an $\mathcal{L}_1$ norm on a vector space. If one uses it to compute all the distances among a group of spike trains elicited from a common stimulus, and then embeds those spike trains using those distances by MDS, the resulting picture is very complicated. The embedded set of spike trains may have hyperbolic structure that is not present in the stimulus space.

We propose a spike metric consistent with the time-coding hypothesis of spike generation that has all of the desirable properties of an $\mathcal{L}_p$ norm. When $p = 1$, this metric is equal to the Victor-Purpura metric \cite{victorpurpura1996}, \cite{victorpurpura1997}. When $p = 2$, embeddings of spike trains using this metric `fit' in Euclidean space with substantially less difficulty than they do for any other value of $p \geq 1$, so advanced dimensionality-reduction and clustering techniques may be used. The dynamic programming algorithm described by Victor and Purpura does not work when $p > 1$, so we propose a faster quadratic-time algorithm that works for all $p \geq 1$.

This procedure we use to calculate our metric is the Hungarian Algorithm \cite{schrijver2003}, which can be made substantially faster for the specific task of comparing spike trains. The crucial insight behind the Hungarian Algorithm was independently discovered by Kuhn and Munkres \cite{kuhn1955}, \cite{munkres1957}. Its function is to solve the minimum-weight matching problem on weighted bipartite graphs. The Hungarian Algorithm is a special case of algorithms to solve general matching problems \cite{papadimitriousteiglitz1998} and assignment problems \cite{burkard1999}. Our version of the algorithm it is a special case of algorithms to solve transportation problems on Monge arrays, described in \cite{burkardklinzrudolf1995}. The original transportation problem was solved by Hoffman \cite{hoffman1963} and formulated by Monge \cite{monge1781}.

\section{The Metric}

We desire to match spikes in one spike train to corresponding spikes in another spike train to compare the two signals. Two trains fired in response to the same stimulus may have different numbers of spikes. So, an alignment of the spikes in the two trains will be inherently imperfect, some spikes may have to be deleted from the trains to make the alignment exact. Victor and Purpura's idea was to break the process of aligning two spike trains into steps, each with their own cost. The metric is equal to the sum of the costs incurred by the most efficient alignment \cite{victorpurpura1996}, \cite{victorpurpura1997}. Let $q > 0$. The cost of aligning two spikes on different trains is $q\Delta t$, where $\Delta t$ is the distance between them. The cost of deleting a spike on either train is $1$. If in two trains the differences between spike times in aligned pairs of spikes are $\Delta t_i$, for $i \in \{1,\ldots, k\}$, the number of spikes deleted from the first train is $D_1$, and the number of spikes deleted from the second train is $D_2$, the distance between the two spike trains is
$$ \sum_{i=1}^{k}q\Delta t_i + D_1 + D_2. $$
The metric is defined to be the minimum of this quantity over all ways of aligning the trains. Our generalization is as follows: Let $p \geq 1$, the metric is
$$ \min\left[\sum_{i=1}^{k}q^p\Delta t_i^p + D_1 + D_2\right]^{1/p}. $$
When $p = 1$ this is the standard Victor-Purpura metric.\\
\indent We can formulate the metric in a way that is more mathematically specific as follows: Let $\textbf{x} = \{x_i\}_{i=1}^{m}$ and $\textbf{y} = \{y_j\}_{j=1}^{n}$ be spike trains, or strictly increasing finite sequences of real numbers. Let $M$ be a `matching' between the two spike trains, a set of ordered pairs $\{(x_{i_1},y_{j_1}),\ldots,(x_{i_k},y_{j_k})\}$ with no element of $\textbf{x}$ or $\textbf{y}$ repeated. Let $\mathcal{M}$ be the set of all matchings. Let the \textit{cardinality} of $M$, $|M|$, be the number of ordered pairs contained in $M$. Define
\begin{equation}
d_{p,q}[M](\textbf{x},\textbf{y}) = \left[ \sum_{(x_i,y_j)\in M}q^p|x_i-y_j|^p + (m-|M|)+(n-|M|) \right]^{1/p}.
\end{equation}
Our metric is
$$ d_{p,q}(\textbf{x},\textbf{y}) = \min_{m\in\mathcal{M}}d_{p,q}[M](\textbf{x},\textbf{y}). $$
We shall show below that we can compute this expression through the Hungarian Algorithm \cite{schrijver2003}, with additional speedups, from $o(mn(m+n)^2)$ time to $o(mn)$ time.

\section{Outline of the algorithm}

Our metric is a minimization over all possible matchings. We can map this minimization to a classical problem, {\em weighted bipartite matching }. In this problem we have a graph whose nodes belong to two different classes with no edge connecting members of the same class (bipartite) and each edge of the graph has a ``weight'' or cost.  A matching or pairing is a subgraph in which each node has at most one connection, and the matching with the smallest total cost is sought. 

A priori, the number of  distinct matchings is enormous. There is a profound result, called the Hungarian Algorithm \cite{schrijver2003}, that shows that only a small subset of these matchings needs to be examined. This result works by showing that the optimal match with precisely $k+1$ pairings (cardinality $k+1$) is related in a rather specific way to the optimal match of cardinality $k$: the optimal $k+1$ match differs from the optimal $k$ match by an ``$M$-augmenting path''. Since there are many fewer such paths than matchings of cardinality $k+1$ the search is considerably faster. 

We shall show that the convexity of our metric translates into a specific property of the graph weights called the Monge property \cite{burkardklinzrudolf1995}. Using the Monge property we can discard large fractions of the $M$-augmenting paths as candidates: the paths have to be ``monotonic'' and cannot ``jump''. We call the resulting, much smaller set of ``$M$-augmenting incompressible monotonic paths'' the {\em shifts}. There are far fewer shifts to explore in connecting level $k$ with level $k+1$, at most $m+n-2k-1$. 

\section{The Shift Algorithm}

For certain matchings $M$ we can construct a unique two-row matrix representation, or `matching matrix' as follows: The first row contains the elements of $\textbf{x}$ in increasing order (from left to right), and the second row contains the elements of $\textbf{y}$ in increasing order. If $(x_i,y_j)\in M$, then $x_i$ and $y_j$ are in the same column of the matrix. If $x_i$ is not matched in $M$, then it is in the same column as a `$*$' symbol. The same goes for unmatched elements of $\textbf{y}$. Any unmatched elements of $\textbf{x}$ and $\textbf{y}$ together appear in increasing order from left to right in $M$. Making this matrix takes $o(m+n)$ time. For instance, let $\textbf{x} = \{b,e,f,h\}$, $\textbf{y} = \{a,c,d,g,i\}$, where $a < b < \cdots < h < i$, with $M = \emptyset$. The matching matrix looks like:
$$ \left[\begin{array}{ccccccccc}
* & b & * & * & e & f & * & h & * \\
a & * & c & d & * & * & g & * & i
\end{array}\right] $$
Now assume `$b$' and `$h$' from $\textbf{x}$ are matched with `$c$' and `$g$' from $\textbf{y}$. The new matrix looks like:
$$
\left[\begin{array}{ccccccc}
*&b&*&e&f&h&*\\
a&c&d&*&*&g&i
\end{array}
\right]
$$
Define the `shift' operation on a matching matrix as follows: Delete two $*$'s, one on the top and one on the bottom and such that there exist no other $*$'s on either row between them. Then shorten the rows to compensate. Below are both possible shifts applied to the matrix above:
$$
\left[\begin{array}{cccccc}
*&b&e&f&h&*\\
a&c&d&*&g&i
\end{array}
\right],
\left[\begin{array}{cccccc}
*&b&*&e&f&h\\
a&c&d&*&g&i
\end{array}
\right]
$$
\textbf{Main Theorem.} \textit{If $M$ minimizes $d_{p,q}[M](\textbf{x},\textbf{y})$ over matchings of cardinality $k < \min(m,n)$, then at least one matching that minimizes $d_{p,q}[M](\textbf{x},\textbf{y})$ over matchings of cardinality $k+1$ is some shift of the matching matrix of $M$.}\vskip .06 in

The proof is outlined in Section 6 and completed in the Appendix. The empty matching always has a matching matrix, and any shift applied to a matching matrix returns a matching matrix. Therefore, we can find the optimal matching of cardinality $k+1$ by tabulating all shifts on the optimal matching of cardinality $k$ and then picking the one that minimizes $d_{p,q}[M](\textbf{x},\textbf{y})$. Even if there are two optimal matchings, $M_1$ and $M_2$ of cardinality $k+1$, and only $M_1$ is a shift of the matrix corresponding to the optimal matching of cardinality $k$, at least one optimal matching of cardinality $k+2$ is a shift of $M_1$.

The time cost of the naive implementation of the algorithm is as follows: Making the matching matrix takes $o(m+n)$ time. Finding the distance-minimizing matching of order $k+1$ by calculating the expected gain or loss to the distance function from every possible shift on the distance-minimizing matching of order $k$ takes $o(m+n)$ time. There are $o(\min(m,n))$ rounds of searching, so we get $o((m+n)\min(m,n))$ time.

We have implemented two key speed-ups. $\min(m,n)$ rounds of shifting is the worst-case scenario; by Theorem 3 the algorithm can terminate as soon as all possible shifts increase the value of $d_{p,q}[M](\textbf{x},\textbf{y})$. After making the first matching matrix, cut it in two between any two spikes (on either train) separated by more than $2^{1/p}/q$. By Corollary 4, the optimal matching is the union of the two optimal matchings found by repeatedly shifting these two submatrices. To prove the algorithm's correctness we need a result from graph theory.

\section{The Hungarian Algorithm}

For the sake of completeness we condense Chapter 3 of \cite{schrijver2003} below. Here we demonstrate one way in which the optimal matching of cardinality $k+1$ is related to the optimal matching of cardinality $k$. This is the first step toward showing that the relationship is in fact a shift.\vskip .3 cm
\noindent\textbf{Definitions.} A graph $G = (V,E)$ is a set of vertices $V$ and edges $E$, where an edge $e$ is defined to be a pair of two vertices, $e = \{v,v'\}$, where $v,v'\in V$. $G$ is complete bipartite if $V = A\cup B$ where $A\cap B = \emptyset$ and every edge $e$ contains exactly one element of each of $A$ and $B$. $G$ is edge-weighted if there exists a weight function $w:E\longrightarrow\mathbb{R}$ on the edges of $G$. From now on all graphs mentioned will be complete bipartite and edge-weighted. A matching $M$ on $G$ is a subset of $E$ such that no two edges in $M$ share a vertex. Let $w(M) = \sum_{e\in M}w(e)$ be the weight of matching $M$.\\
\indent A path $P$ on $G$ is an ordered subset of distinct elements of $E$, $(e_1,\ldots,e_t)$, such that $e_i$ and $e_{i+1}$ share a vertex for all $1\leq i < t$ but $e_1$ and $e_t$ do not share a vertex. A path $P$ is $M$-augmenting for a matching $M$ if: 1. $t$ is odd. 2. $e_2,e_4,\ldots,e_{t-1}\in M$. 3. $e_1$ and $e_t$ each only share one vertex with an element of $M$. For an example of an $M$-augmenting path, see Figure 1. For an edge $e$, define its length function $l(e)$ to be $w(e)$ if $e\in M$ and $-w(e)$ if $e\notin M$. For a path $P$, let $l(P) = \sum_{e\in P}l(e)$.\\
\indent If $C,D\subseteq E$, $C\triangle D$ contains all edges that are in exactly one of $C$ and $D$. It follows that if $P$ is $M$-augmenting then $M\triangle P$ is a matching of cardinality $|M|+1$. \vskip .06 in
\noindent\textbf{Theorem 1.} \textit{\emph{\cite{schrijver2003}, Section 3.5, Proposition 1}}: Let $P$ be an $M$-augmenting path of maximum length. If $M$ is a matching of minimum weight of cardinality $k$, then $M' = M\triangle P$ is a matching of minimum weight of cardinality $k+1$.\vskip .06 in
\noindent The proof is in the appendix.

\section{Proof of Shift Algorithm}

Here we outline the proof that augmenting paths of maximum length correspond to shifts. The details of the proof are in the Appendix.

\noindent\textbf{Lemma 1.} \textit{Computing $d_{p,q}(\textbf{x},\textbf{y})$ is equivalent to finding a minimum weight matching on the complete bipartite weighted graph with vertex sets $A = \{a_1,\ldots,a_m\}$ and $B=\{b_1,\ldots,b_n\}$ and weight function $w(a_i,b_j) = q^p|x_i-y_j|^p - 2$.}\vskip .06 in
\begin{proof}
Every matching $M$ on this graph is assocated with a matching $M$ (abusing notation) between spike trains $x$ and $y$. To minimize $d_{p,q}[M](\textbf{x},\textbf{y})$ over $M$ it is equivalent to minimize
$$ \sum_{(x_i,y_j)\in M}\left(q^p|x_i-y_j|^p - 2\right) = \sum_{(a_i,b_j)\in M}w(a_i,b_j) = w(M) $$
\end{proof}

Now we need to find an analogue for shifting a matching matrix in terms of performing an operation on graphs. It turns out that the analogue is taking the symmetric difference between $M$ and a special type of $M$-augmenting path, which we will also call a ``shift''. It is a path of the form:
$$(\{a_{i},b_{j}\},\{a_{i},b_{j+1}\},\{a_{i+1},b_{j+1}\}\ldots,\{a_{i+N},b_{j_{j+N}}\})$$
or
$$(\{a_{i},b_{j}\},\{a_{i+1},b_{j}\},\{a_{i+1},b_{j+1}\},\ldots,\{a_{i+N},b_{j+N}\}),$$
with a couple additional constraints: in the first case there are no unmatched elements of $\textbf{x}$ between $x_i$ and $y_j$ and there are no unmatched elements of $\textbf{y}$ between $x_{i+N}$ and $y_{i+N}$, and in the second case there are no unmatched elements of $\textbf{y}$ between $x_i$ and $y_j$ and there are no unmatched elements of $\textbf{x}$ between $x_{i+N}$ and $y_{i+N}$. See Figure 2 for an $M$-augmenting path that is a shift. In this figure and all subsequent figures we place the sets of nodes corresponding to the spikes on trains $\textbf{x}$ and $\textbf{y}$ on parallel lines, and nodes farther to the right correspond to spikes occurring later in time than the spikes associated to nodes on their left. We can now restate the Main Theorem:\vskip .06 in

\noindent\textbf{Main Theorem, Restated.} \textit{Let a bipartite graph have vertex sets $A = \{a_1,\ldots,a_m\}$ and $B=\{b_1,\ldots,b_n\}$ and weight function $w(a_i,b_j) = q^p|x_i-y_j|^p - 2$. Let $M$ minimize $w(M)$ over matchings of cardinality $k < \min(m,n)$. Then there exists an $M$-augmenting path $P$ that is a shift such that $M\triangle P$ minimizes $w(M)$ over matchings of cardinality $k+1$.}\vskip .06 in

\noindent\textbf{Lemma 2.} \textit{\emph{Strict Monge Property:}} 
Let $x_{i_1} < x_{i_2}$ and $y_{j_1} < y_{j_2}$. For $p > 1$,
$$ |x_{i_1} - y_{j_1}|^p + |x_{i_2} - y_{j_2}|^p < |x_{i_1} - y_{j_2}|^p + |x_{i_2} - y_{j_1}|^p$$\vskip .06 in
The proof uses convexity and is in the Appendix.\vskip .06 in

\noindent\textbf{Lemma 3.} \textit{Let $p > 1$. If $M$ is a minimum weight matching of cardinality $k$ containing edges $\{a_{i_1},b_{j_1}\}$ and $\{a_{i_2},b_{j_2}\}$ then either $i_1<i_2$ and $j_1<j_2$ or $i_1>i_2$ and $j_1>j_2$.}\vskip .06 in
\begin{proof} Remove edges $\{a_{i_1},b_{j_1}\}$ and $\{a_{i_2},b_{j_2}\}$ and replace them with edges $\{a_{i_1},b_{j_2}\}$ and $\{a_{i_2},b_{j_1}\}$ to form the matching $M'$ of cardinality $k$. By Lemma 2,
$$ w(M')-w(M) = q^p|x_{i_1}-y_{j_2}|^p+q^p|x_{i_2}-y_{j_1}|^p-q^p|x_{i_1}-y_{j_1}|^p-q^p|x_{i_2}-y_{j_2}|^p $$
Since $M$ is of minimum weight, this expression must be positive. Lemma 2 implies that either $i_1<i_2$ and $j_1<j_2$ or $i_1>i_2$ and $j_1>j_2$.
\end{proof}
This lemma can be pictorially understood as meaning if $M$ is a minimum weight matching of cardinality $k$, it contains no edges that cross, using our convention for drawing figures stated below Lemma 1. See Figure 3 for a demonstration of Lemma 3.\vskip .06 in

\noindent\textbf{Theorem 2.} \textit{Let $p > 1$, let $M$ be a minimum weight matching of cardinality $k$, and let $P$ be an $M$-augmenting path of maximum length. Then $P$ is of the form
$$(\{a_{i_1},b_{j_1}\},\{a_{i_1},b_{j_2}\},\{a_{i_2},b_{j_2}\},\ldots,\{a_{i_l},b_{j_{l}}\})$$
or
$$(\{a_{i_1},b_{j_1}\},\{a_{i_2},b_{j_1}\},\{a_{i_2},b_{j_2}\},\ldots,\{a_{i_{l}},b_{j_l}\})$$
where the sequences $\{a_{i_t}\}$ and $\{b_{i_t}\}$ are either strictly increasing or strictly decreasing.}\vskip .06 in
The exact proof is in the Appendix. It can be understood using diagrams. What we want to show is that $P$ does not turn back on itself, it either moves forwards or backwards through the graph. If it should decide to change direction, one of two things happen. It might make use of two edges in $M$ that cross, contradicting Lemma 3, see Figure 4a. If not, it uses two edges that are both not in $M$ that cross (if it uses an edge in $M$ and an edge not in $M$ that cross, one of the previous two problems necessarily occur). In the second case, both of those edges will be in $M\triangle P$. But if $P$ is of maximum length, by Theorem 1, $M\triangle P$ is a minimum matching of cardinality $k+1$, which also cannot contain crossing edges by Lemma 3, see Figure 4b. This is a contradiction, and it follows that $P$ cannot turn around.\\

We are now ready to sketch the proof of the Main Theorem for $p > 1$. It follows for $p = 1$ by real analysis. Both proofs are in the Appendix.

Let $M$ be a minimum weight maching of cardinality $k$, and $P$ be an $M$-augmenting path of maximum length. $P$ is of the form described by Theorem 2. Assume without loss of generality that the vertex at which $P$ starts is $a_{i'}$ and the vertex at which it ends is $b_{j'}$. There are two things that could go wrong to prevent transforming $M$ into $M\triangle P$ from being a shift on $M$'s matching matrix. One is that there exists an unmatched spike on either spike train between $x_{i'}$ and $y_{j'}$. In that case, it can be shown that if $M$ is the minimum matching of cardinality $k$, then $M\triangle P$ is not the minimum matching of cardinality $k+1$; it can be improved upon by replacing an edge. See Figures 5a and 5b. The second possible problem is that there exists an edge in $M$ that some edge in $P$ crosses. Then, it can be shown that $M$ or $M\triangle P$ contain crossing edges, contradicting Lemma 3. See Figures 6a and 6b. With these two possibilities eliminated, $P$ could only be an $M$-augmenting path corresponding to a shift.\vskip .06 in

\noindent\textbf{Theorem 3.} \textit{Let $M$ be a minimum weight matching of cardinality $k$. If a minimum weight matching of cardinality $k+1$ has greater weight then $M$, then $M$ is a minimum weight matching over all matchings of any cardinality.}\vskip .06 in
\noindent The proof is in the Appendix.\vskip .06 in

\noindent\textbf{Corollary 4.} \textit{Let $M_k$ denote the minimum weight matching of cardinality $k$. If $M_{k_0}$ is the minimum weight matching over matchings of all cardinalities, than no $M_k$ contains a positive-weighted edge for $k \leq k_0$.} \vskip .06 in
\begin{proof}
The proposition is vacuous if $k_0 = 0$. So assume the statement is true for $k$, $0 < k < k_0$, we will prove it for $k+1$. Assume that $M_{k+1}$ has an edge $e$ of positive weight. Then $M_{k+1}\triangle\{e\}$ is a matching of cardinality $k$, so $w(M_{k+1}\triangle\{e\}) \geq w(M_k)$. But $w(M_{k+1}\triangle\{e\}) < w(M_{k+1}) \leq w(M_k)$ by Theorem 3, a contradiction.
\end{proof}
The implication of this corollary is that if any two adjacent columns of a matching matrix contain entries that are separated by more than $2^{1/p}/q$, the matrix can be cut in two and the two halves optimized separately.

\section{Conclusion}

We have presented a spike metric satisfying two important desiderata: That it be grounded in the time-coding hypothesis of spike generation, and that it be closely related to the Euclidean $\mathcal{L}_2$ norm. The latter property is important to subsequent stages of analysis of spike data, especially if MDS is used.

We have also presented a fast algorithm to calculate our metric and proved its correctness and amortized complexity. We demonstrated that it is faster than existing algorithms to compute the $p = 1$ special case. Our proof used the toolboxes of graph theory and combinatorial optimization, and demonstrates that they can be usefully brought to bear on problems in computational biology.

We conjecture that related procedures to the one we have presented could be used to align both spike trains with inputs from multiple neurons (as done in \cite{aronov2003}) and strands of DNA.

Future work will include applications to multiple alignment of neural signals and DNA. Our metric suggests new measures of the quality of multiple alignments, and our algorithm suggests new ways of computing them. One approach to aligning multiple spike trains would be to generalize the CLUSTAL procedures for `progressive' alignment, i.e. building a multiple alignment from a collection of aligned pairs \cite{higginssharp1988}, \cite{thompsonhigginsgibson1994}.

\section{Appendix}

\noindent\textbf{Definitions.} $G$ is connected if one can travel from any vertex to any other vertex via the edges. A component of $G$ is a maximal connected subgraph.
A path in which the first and last edges share a vertex is called a circuit.\\

\noindent\textbf{Lemma 4.} \textit{\emph{\cite{schrijver2003}, Theorem 1:} If $M$ is a matching in $G$ then either there exists an $M$-augmenting path $P$ or no matching of greater cardinality exists. If $N$ is a matching of greater cardinality, we can pick $P \subseteq M\cup N$.}\vskip .06 in
\noindent\textit{Proof.} If $M$ is a matching of maximum cardinality and $P$ is an $M$-augmenting path then $M\triangle P$ would be a matching of greater cardinality, a contradiction. If $M$ is not a matching of maximum cardinality, there exists a larger matching $N$. Let $H$ be the graph with vertices $V$ and edges $M\cup N$. Every connected component of $H$ is a path or a circuit. As $|N| > |M|$ one of these components $C$ contains more edges of $N$ than $M$. Its edges must alternate between those in $M$ and those in $N$, since no two edges of either one can be connected to the same vertex. Thus, in order to have more edges in $N$ than in $M$, it must start and end with non-identical edges in $N$. Therefore, $C$ is an $M$-augmenting path.\vskip .06 in

\noindent\textit{Proof of Theorem 1.} From \cite{schrijver2003}, Section 3.5, Proposition 1: Let $N$ be an arbitrary matching of size $k+1$. By Lemma 4 we can pick an $M$-augmenting path $Q$ in $M\cup N$. By definition $l(Q)\leq l(P)$. $|N\triangle Q| = k$, and since $M$ is of minimum weight for cardinality $k$, $w(N\triangle Q)\geq w(M)$. Therefore,
$$ w(N) = w(N\triangle Q)-l(Q)\geq w(M)-l(P) = w(M') $$
Therefore $M'$ is of minimum weight over matchings of cardinality $k+1$.\vskip .06 in

\noindent\textit{Proof of Lemma 2.}
Let $\alpha = x_{i_1}-y_{j_2}$, $\beta = x_{i_2}-y_{j_1}$, and $\gamma = y_{j_2}-y_{j_1}$. Now, $0 < \gamma < \beta-\alpha$. We want
$$ f(\gamma) = |\alpha|^p + |\beta|^p - |\alpha+\gamma|^p - |\beta-\gamma|^p > 0. $$
The left hand side has roots for $\gamma$ at $0$ and $\beta-\alpha$.
$$ f'(\gamma) = -p\cdot \text{sign}(\alpha+\gamma)|\alpha+\gamma|^{p-1}+p\cdot \text{sign}(\beta-\gamma)|\beta-\gamma|^{p-1}. $$
This expression is zero if and only if $\gamma = (\beta-\alpha)/2$. By Jensen's Inequality,
$$ f\left(\frac{\beta-\alpha}{2}\right) = |\alpha|^p+|\beta|^p-2\left|\frac{\alpha+\beta}{2}\right|^p > 0 $$
The lemma follows by Rolle's Theorem.\vskip .06 in

\noindent\textit{Proof of Theorem 2.} We assume without loss of generality that $P$ is of the form 
$$(\{a_{i_1},b_{j_1}\},\{a_{i_1},b_{j_2}\},\ldots,\{a_{i_l},b_{j_{l}}\}).$$
Assume for the sake of contradiction that for some $t$ either $i_t < i_{t+1}$ and $i_{t+1} > i_{t+2}$ or $i_t > i_{t+1}$ and $i_{t+1} < i_{t+2}$ (The cases where for some $t$ either $j_t < j_{t+1}$ and $j_{t+1} > j_{t+2}$ or $j_t > j_{t+1}$ and $j_{t+1} < j_{t+2}$ are analagous). The cases are similar so we only treat the first one. It has two subcases: $j_{t+1} > j_{t+2}$ and $j_{t+1} < j_{t+2}$. In the first subcase, the edges $\{a_{i_{t}},b_{j_{t+1}}\}$ and $\{a_{i_{t+1}},b_{j_{t+2}}\}$ are both in $M$, contradicting Lemma 3. In the second subcase case, the edges $\{a_{i_{t+1}},b_{j_{t+1}}\}$ and $\{a_{i_{t+2}},b_{j_{t+2}}\}$ are both in $M\triangle P$, which by Theorem 1 is a minimum weight matching of cardinality $k+1$, contradicting Lemma 3.\vskip .06 in

\noindent\textit{Proof of Main Theorem.} We show that if $M$ is a minimum weight matching of cardinality $k$ then an $M$-augmenting path of maximum length, $P$, is of the form 
$$(\{a_{i},b_{j}\},\{a_{i},b_{j+1}\},\ldots,\{a_{i+N},b_{j+N}\}) \text{ or } (\{a_{i},b_{j}\},\{a_{i+1},b_{j}\},\ldots,\{a_{i+N},b_{j+N}\})$$
where in the first case there are no unmatched elements of $\textbf{x}$ between $x_i$ and $y_j$ and there are no unmatched elements of $\textbf{y}$ between $x_{i+N}$ and $y_{i+N}$, and in the second case there are no unmatched elements of $\textbf{y}$ between $x_i$ and $y_j$ and there are no unmatched elements of $\textbf{x}$ between $x_{i+N}$ and $y_{i+N}$. Therefore, transforming $M$ into $M\triangle P$ is equivalent to the shift operation described in Section 4.\\
\indent We first prove the case where $p > 1$. Using Theorem 2, assume without loss of generality that $P$ is of the form
$$(\{a_{i_1},b_{j_1}\},\{a_{i_1},b_{j_2}\},\ldots,\{a_{i_l},b_{j_{l}}\})$$
where $\{a_{i_t}\}$ and $\{b_{i_t}\}$ are strictly increasing. The theorem is clear when $l = 1$. So let $l > 1$ and assume that there exists and integer $r$ such that $i_t < r < i_{t+1}$ (we can likewise assume that there exists an integer between $j_t$ and $j_{t+1}$, this case is similar). Now there are two cases: either $a_{r}$ is connected to an edge in $M$ or it is not. If it is connected to an edge, call that edge $\{a_{r},b_s\}$. $s > j_{t+1}$, otherwise the fact that $\{a_{r},b_s\}$ and $\{a_{i_{t}},b_{j_{t+1}}\}$ are both in $M$ contradicts Lemma 3. But in that case, since $\{a_{r},b_s\}\notin P$, $\{a_{r},b_s\}$ and $\{a_{i_{t+1}},b_{j_{t+1}}\}$ are both in $M\triangle P$, which by Theorem 1 is a minimum weight matching of cardinality $k+1$, contradicting Lemma 3.\\
\indent Now consider the case when $a_{r}$ is not connected to an edge. Replace the edge $\{a_{i_{t+1}},b_{j_{t+1}}\}$ with $\{a_{r},b_{j_{t+1}}\}$ in $P$ to form $P'$. $P'$ may not be an $M$-augmenting path, but $M\triangle P'$ is still a matching of cardinality $k+1$. We know that $q^p|x_{i_t}-y_{j_{t+1}}|^p \leq q^p|x_{r}-y_{j_{t+1}}|^p$ since otherwise the edge $\{a_{r},b_{j_{t+1}}\}$ would be in $M$ instead of $\{a_{i_t},b_{j_{t+1}}\}$. Therefore, $|x_{i_t}-y_{j_{t+1}}|\leq|x_{r}-y_{j_{t+1}}|$. If $y_{j_{t+1}}\geq x_{r}$, this implies $-x_{i_t}+y_{j_{t+1}}\leq -x_{r}+y_{j_{t+1}}$ making $x_{i_t} \geq x_{r}$, a contradiction. So $x_{r} > y_{j_{t+1}}$. Since $x_{i_{t+1}} > x_{r}$, $x_{r}-y_{j_{t+1}} < x_{i_{t+1}}-y_{j_{t+1}}$, making
$$ w(M\triangle P')-w(M\triangle P) = q^p|x_{r}-y_{j_{t+1}}|^p - q^p|x_{i_{t+1}}-y_{j_{t+1}}|^p < 0. $$
$M\triangle P'$ is a matching of cardinality $k+1$ of lower weight than that of $M\triangle P$, but since $P$ is of maximum length, Theorem 1 implies that $M\triangle P$ has minimum weight. This is a contradiction and the theorem follows for $p > 1$.\\
\indent Finally, assume that there is an unmatched element of $\textbf{x}$ between $x_{i_1}$ and $y_{j_1}$, call it $x_u$, with corresponding vertex $a_u$ (the case when there is an unmatched element of $\textbf{y}$ between $x_{i_l}$ and $y_{j_l}$ is analogous). Since $x_u < x_{i_1}$ by the previous paragraph, $y_{j_1} \leq x_u < x_{i_1}$. Replace the edge $\{a_{i_1},b_{j_1}\}$ with $\{a_u,b_{j_1}\}$ in $P$ to form $P'$. $P'$ may be be an $M$-augmenting path, but as before $M\triangle P'$ is still a matching of cardinality $k+1$. $x_u < x_{i_1}$ implies $x_u - y_{j_1} < x_{i_1} - y_{j_1}$, and since these quantities are nonnegative,
$$ w(M\triangle P')-w(M\triangle P) = q^p|x_u-y_{j_1}|^p - q^p|x_{i_1}-y_{j_1}|^p < 0 .$$
Using Theorem 1, $P$ could not have been an $M$-augmenting path of maximum length, a contradiction. The theorem follows for $p>1$.\\
\indent Now we address the $p = 1$ case. For some constants $C\in\mathbb{Z}$, $\alpha_1,\ldots,\alpha_J>0$, and $u_1,\ldots,u_J\in\{\pm 1\}$, the difference in length between two $M$-augmenting paths $P$ and $P'$ is
$$ g(p) = l(P')-l(P) = q^{p}\sum_{i=1}^{J}u_i\alpha_i^{p} + 2C. $$
$g(p)$ is real analytic and its Taylor series $T(p)$ converges everywhere. Either $g(p)$ is identically $0$ or there exist some least integral exponent $e$ such that the coefficient of $p^e$ in $T(p)$ is nonzero. If it is positive then there exists some $\epsilon_1>0$ such that $f(p)$ is strictly positive on $(1,1+\epsilon_1)$, likewise if it is negative we can pick $\epsilon_1$ such that $g(p)$ is strictly negative on $(1,1+\epsilon_1)$ by \cite{nash1959}, Theorem 1. We can determine such an $\epsilon_1$ for every pair of $M$-augmenting paths $P$ and $P'$ that do not have the same length for all $p$. Let $\epsilon_2$ be the minimum of all of these $\epsilon_1$'s. There exists a set of $M$-augmenting paths $Q_1,\ldots,Q_K$ such that for $p\in(1,\epsilon_2)$ they all have equal length and they are all longer than all other $M$-augmenting paths. They all must be shifts. By continuity, when $p = 1$ they still have equal length and are at least as long as all other $M$-augmenting paths. Therefore, $M\triangle Q_1$ is a matching of minimum weight of cardinality $k+1$ when $p=1$.\vskip .06 in

\noindent\textbf{Definition.} Let $M$ be a matching of minimum weight of cardinality $k$ and $P$ be an $M$-augmenting path. We call $P$ an $M$-shift it it corresponds to a shift in the matching matrix of $M$.\vskip .06 in

\noindent\textit{Proof of Theorem 3.} Let $M_k$ be the minimum weight matching of cardinality $k$. It suffices to show that if $w(M_k) < w(M_{k+1})$, then if $M_{k+2}$ exists, $w(M_{k+1}) < w(M_{k+2})$. It follows by induction that $w(M_k)$ has lower weight than all minimum weight matchings of cardinality greater than $k$. Therefore, the sequence $w(M_1),w(M_2),\ldots,w(M_{\min(m,n)})$ is decreasing then increasing, and the point at which the switch occurs is its minimum.\\
\indent All $M_k$-shifts have negative length, since the one with the greatest length, $P$, has negative length, as $w(M_k) < w(M_{k+1})$ (using Theorem 1). Therefore all $M_{k+1}$-shifts have negative length since they are also $M_k$ shifts, except for possibly one, $Q$, that is not. If $Q$ does not exist or is not the $M_{k+1}$-shift of greatest length, then the shift of greatest length has negative length, and $w(M_{k+1}) < w(M_{k+2})$, so we are done. So assume that $Q$ is the $M_{k+1}$-shift of greatest length. We shall prove that $l(Q)\leq 0$. Sending $M_k$ to $M_k\triangle P = M_{k+1}$ and then sending $M_{k+1}$ to $M_{k+1}\triangle Q = M_{k+2}$ corresponds to deleting two pairs of stars in one of four classes of matching matrices. We consider only two of them, since the others are analogous:\\
$$ \left[\begin{array}{ccccccccccc}
\cdots & x_i & x_{i+1} & \cdots & *       & \cdots & x_{i+s} & \cdots & x_{i+t}   & *       & \cdots \\
\cdots & *   & y_j     & \cdots & y_{j+r} & \cdots & *       & \cdots & y_{j+t-1} & y_{j+t} & \cdots
\end{array}\right] $$\\
$$ \left[\begin{array}{ccccccccccc}
\cdots & x_i & x_{i+1} & \cdots & x_{i+r} & \cdots & *       & \cdots & x_{i+t}   & *       & \cdots \\
\cdots & *   & y_j     & \cdots & *       & \cdots & y_{j+s} & \cdots & y_{j+t-1} & y_{j+t} & \cdots
\end{array}\right] $$\\
To get the other two cases, switch the top stars to the bottom and the bottom stars to the top in each of the above examples, and reindex accordingly. In both matrices, deleting the inner pair of stars corresponds to transforming $M_k$ into $M_{k+1}$ and deleting the outer pair of stars corresponds to transforming $M_{k+1}$ into $M_{k+2}$. In the first matching matrix there are three possible $M_k$-shifts. The middle one is $P$. Call the rightmost one $P_-$ and the rightmost one $P_+$. $l(P_+) \leq l(P) < 0$, and $l(P_-) \leq l(P)$. $l(P) + l(Q) = l(P_-) + l(P_+)$, since performing the shift $P$ followed by $Q$ is the same operation as performing the shift $P_-$ followed by $P_+$ on the matrix. Therefore $l(Q) = l(P_+) - l(P) + l(P_-) < 0$.\\
\indent Now consider the second matching matrix. There are $M_k$-augmenting paths $P_-$ and $P_+$ such that changing $M_k$ to $M_k\triangle P_-$ or $M_k\triangle P_+$ is equivalent to deleting both the first and third stars in the matrix (from the left) or both the second and fourth stars from the matrix. As before, $l(P_+) \leq l(P) < 0$, $l(P_-) \leq l(P)$, and $l(P) + l(Q) = l(P_-) + l(P_+)$. Again, $l(Q) = l(P_+) - l(P) + l(P_-) < 0$.

\begin{figure}[p]\includegraphics[scale=0.8]{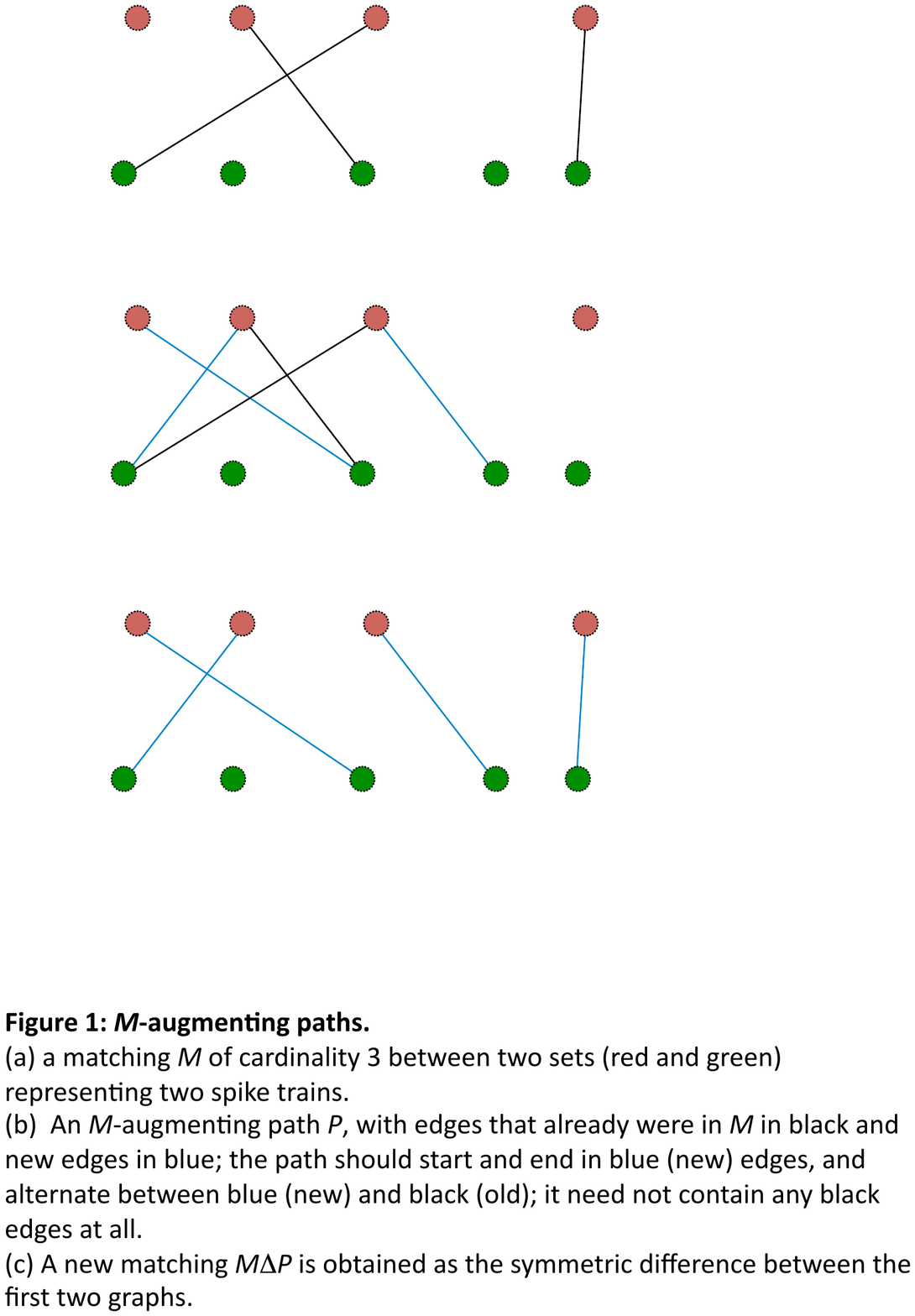}\end{figure}
\begin{figure}[p]\includegraphics[scale=0.8]{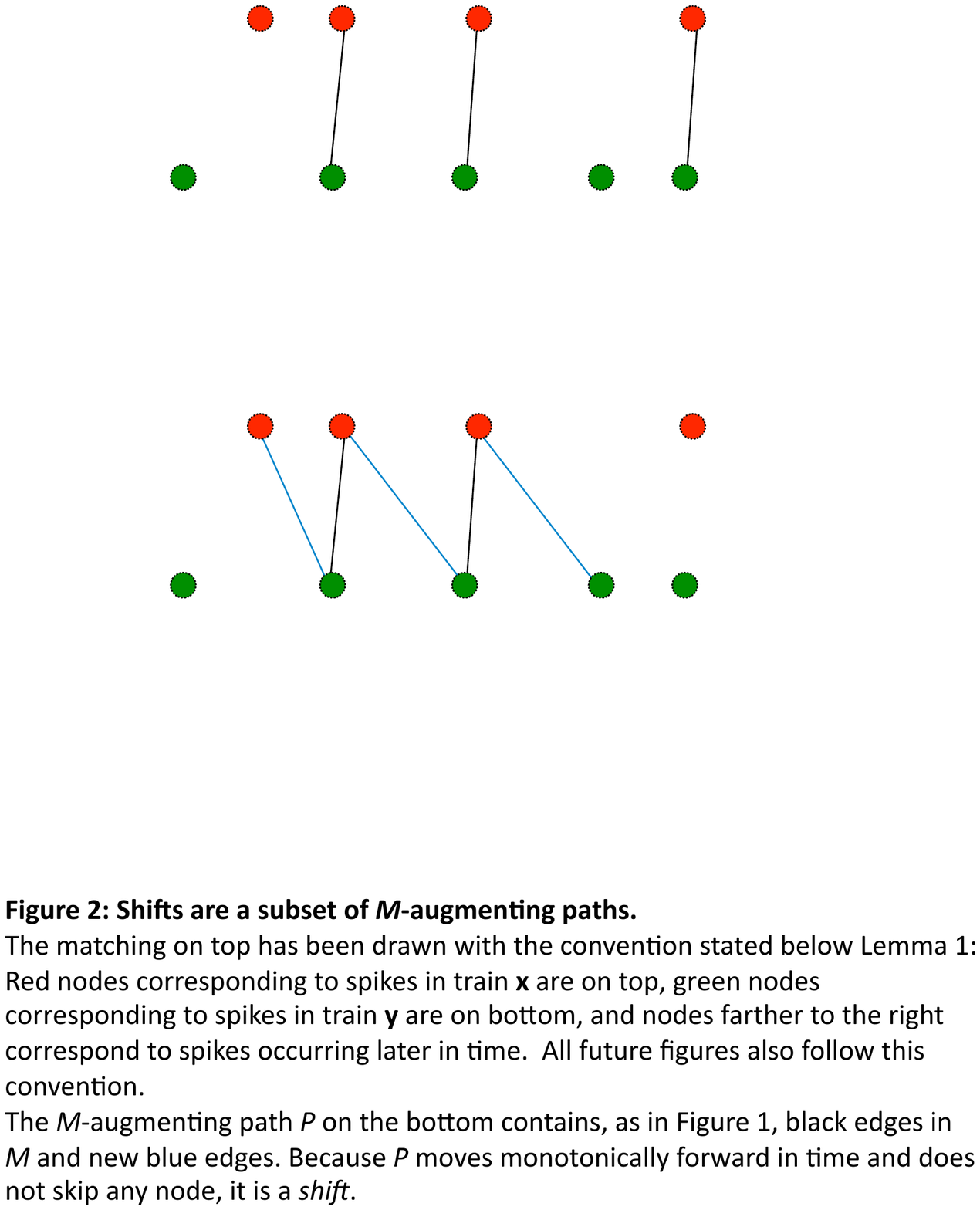}\end{figure}
\begin{figure}[p]\includegraphics[scale=0.8]{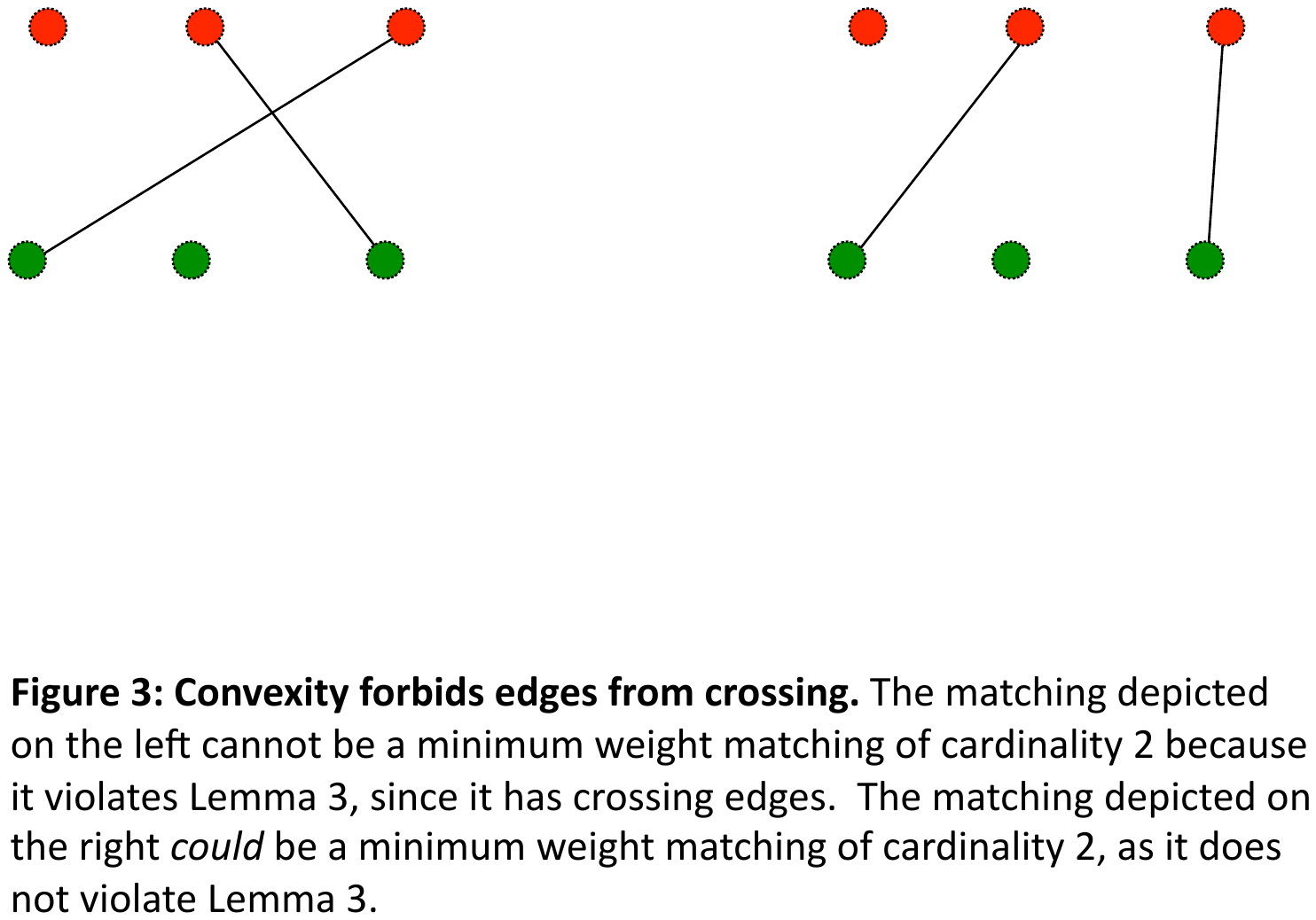}\end{figure}
\begin{figure}[p]\includegraphics[scale=0.8]{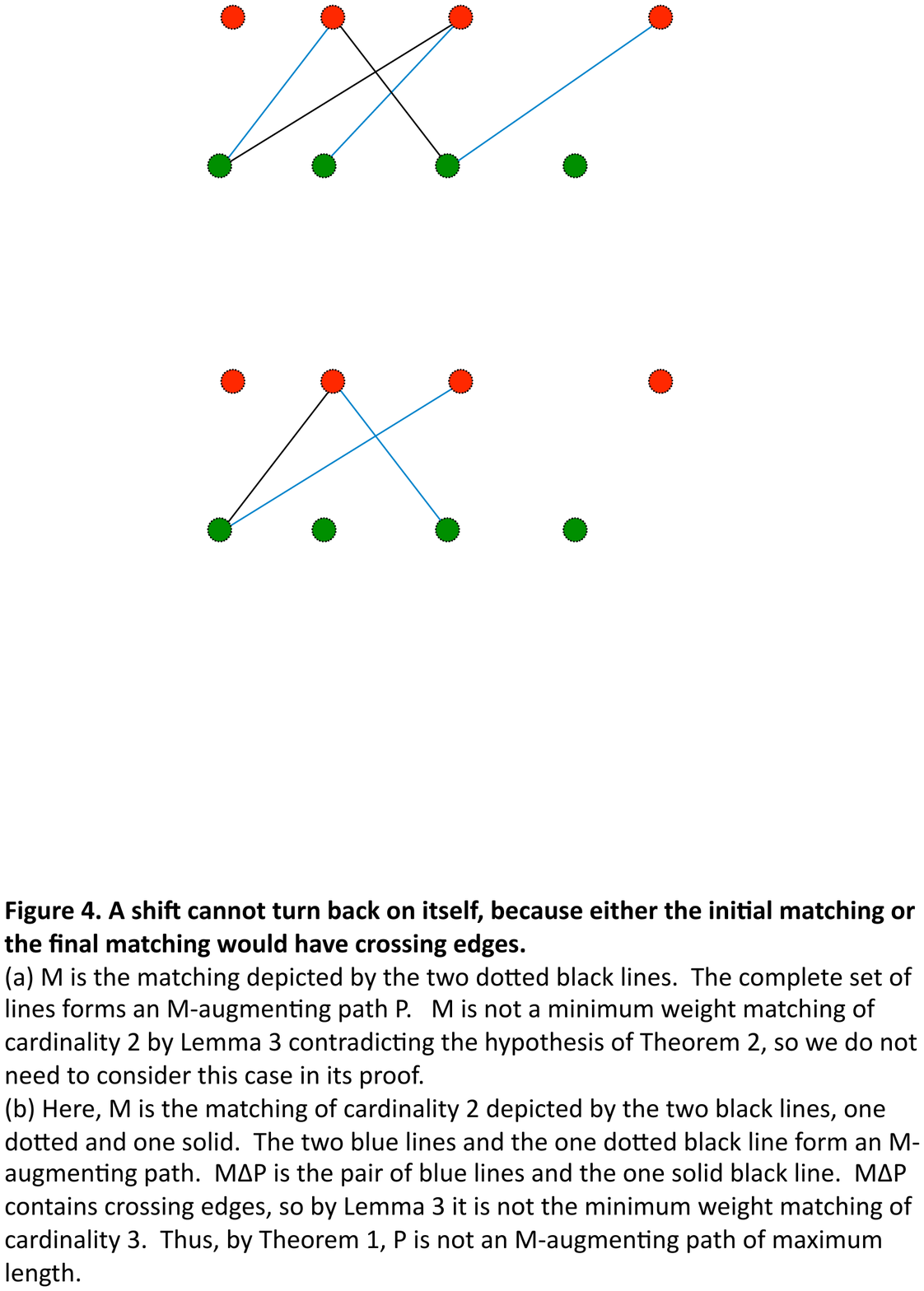}\end{figure}
\begin{figure}[p]\includegraphics[scale=0.8]{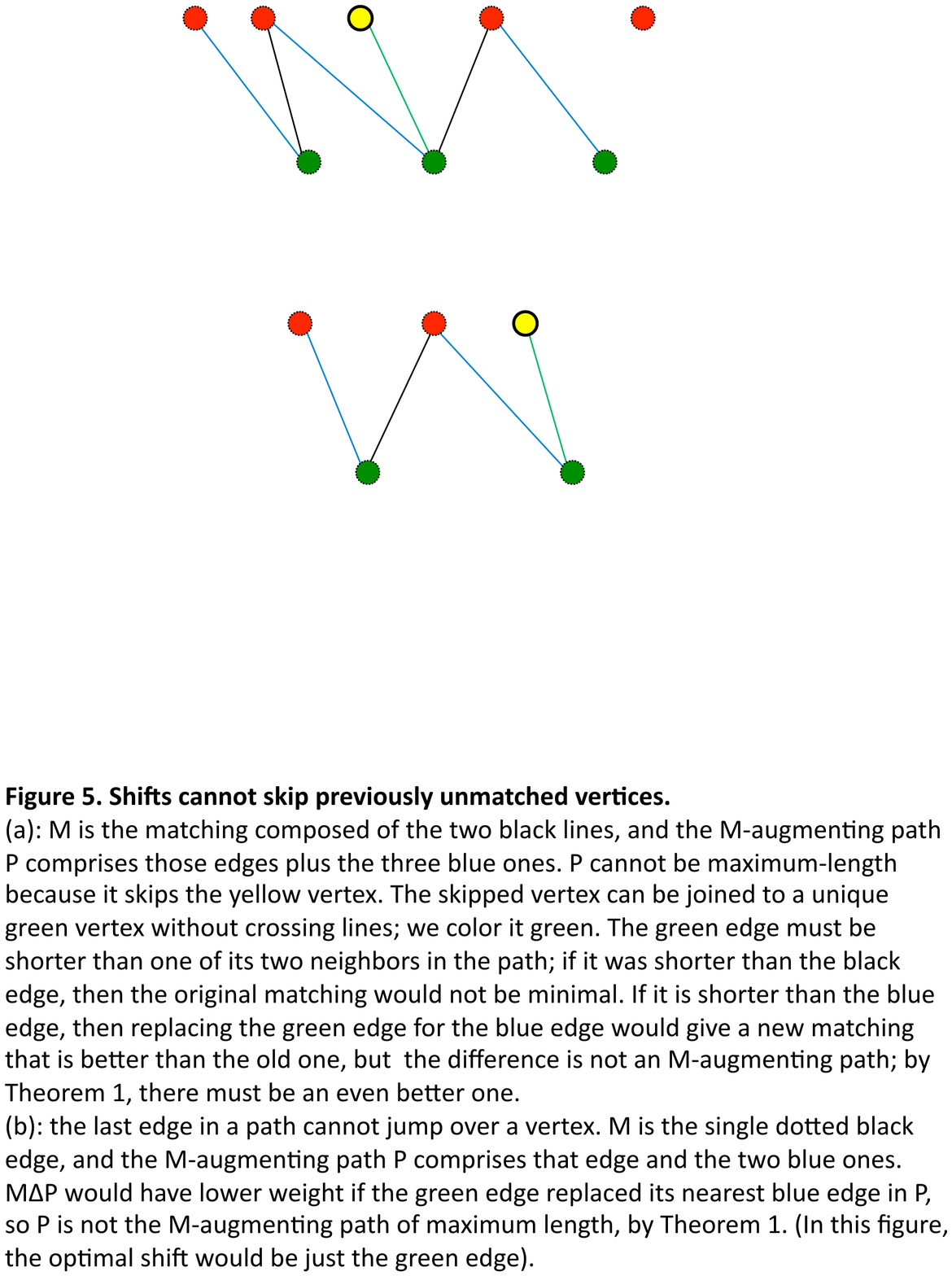}\end{figure}
\begin{figure}[p]\includegraphics[scale=0.8]{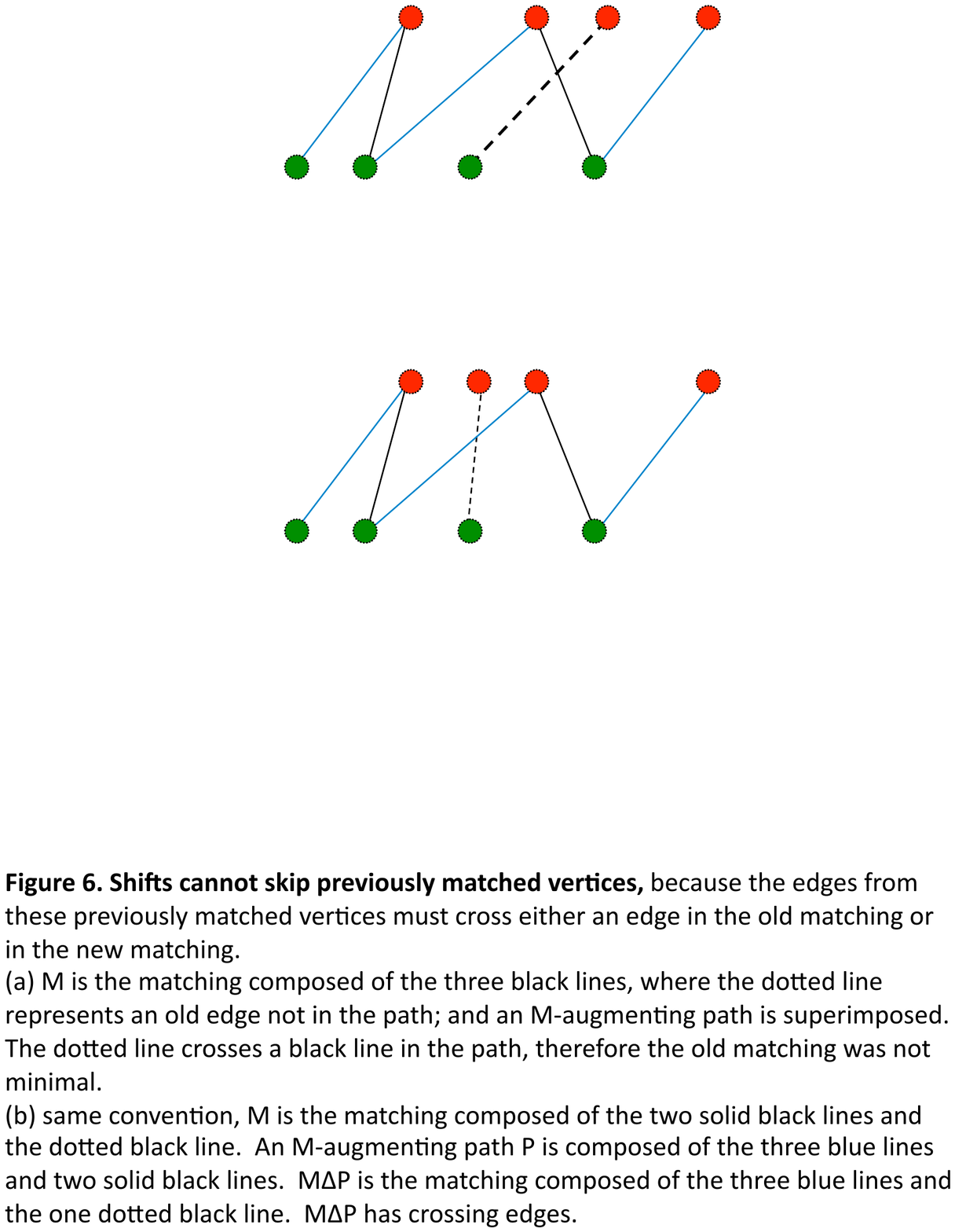}\end{figure}


\begin{thebibliography}{widest-label}

\bibitem[Aronov, 2003]{aronov2003} D. Aronov, ``Fast algorithm for the metric-space analysis of simultaneous responses of multiple single neurons,'' Journal of Neuroscience Methods, Vol 124, Issue 2, (April, 2003) 175-179.

\bibitem[Aronov et al., 2003]{aronovetal2003} D. Aronov, D. Reich, F. Mechler, and J. Victor, ``Neural coding of spatial phase in V1 of the macaque monkey,'' Journal of Neurophysiology 89: 3304-3327, 2003.

\bibitem[Aronov, Victor, 2004]{aronovvictor2004} D. Aronov and J. Victor, ``Non-Euclidean properties of spike train metric spaces,'' Physical Review E 69 (2004) 061905.

\bibitem[Burkard, 1999]{burkard1999} R. Burkard, ``Selected topics on assignment problems,'' Discrete Applied Mathematics, Volume 123, Issue 1-3 (November 2002): 257-302.

\bibitem[Burkard, Klinz, Rudolf, 1995]{burkardklinzrudolf1995} R. Burkard, B. Klinz, and R. Rudolf, ``Perspectives of Monge properties in optimization,'' Discrete Applied Mathematics, Volume 70, Issue 2 (September 1996): 95-161.

\bibitem[Carrillo, Lipman, 1988]{carrillolipman1998} H. Carrillo and D. Lipman, ``The multiple sequence alignment problem in biology,'' SIAM Journal on Applied Mathematics, Vol. 48, No. 5 (Oct, 1988): 1073-1082.

\bibitem[Chase, Young, 2006]{chaseyoung2006} S. Chase and E. Young, ``Spike-timing codes enhance the representation of multiple simultaneous sound-localization cues in the inferior colliculus,'' Journal of Neuroscience, April 12, 2006: 26(15):3889-3898.

\bibitem [Ding, He, 2004]{dinghe2004} C. Ding and X. He, ``K-means Clustering via Principal Component Analysis,'' Proceedings of the International Conference on Machine Learning (ICML 2004), pp 225-232. July 2004.

\bibitem[Gerstner et al., 1997]{gerstneretal1997} W. Gerstner, A. Kreiter, H. Markram, and A. Herz, ``Neural codes: Firing rates and beyond,'' Proceedings of the National Academy of Sciences, Vol. 94, pp. 12740-12741, November 1997.

\bibitem[Hardle, Simar, 2004]{hardlesimar2004} W. Hardle, L. Simar, \textit{Applied Multivariate Statistical Analysis}, Version 26, August 2004, available at: www.quantlet.com/mdstat/scripts/mva/htmlbook/mvahtml.html

\bibitem[Higgins, Sharp, 1988]{higginssharp1988} D. Higgins and P. Sharp, ``CLUSTAL: a package for performing multiple sequence algignment on a microcomputer,'' Gene, 73 (1988): 273-244.

\bibitem[Hoffman, 1963]{hoffman1963} A. Hoffman, ``On simple linear programming problems,'' Proceedings of Symposia in Pure Mathematics Volume VII: Convexity, 1963, pp. 317-327.

\bibitem[Hopfield, 1995]{hopfield1995} J. Hopfield, ``Pattern recognition computation using action potential timing for stimulus representation,'' Nature, Vol. 376, 6, July 1995, pp. 33-36.

\bibitem[Kuhn, 1955]{kuhn1955} H. Kuhn, ``The Hungarian Method for the assignment problem,'' Naval Research Logistics Quarterly, 2:83-97, 1955.

\bibitem[Kruskal, Wish, 1978]{kruskalwish1978} J. Kruskal, M. Wish, \textit{Multidimensional Scaling}, Sage University Paper series on Quantitative Application in the Social Sciences, 07-011. Beverly Hills and London: Sage Publications (1978).

\bibitem[Lipman et al., 1989]{lipmanetal1989} D. Lipman, S. Altschul, and J. Kececioglu, ``A tool for multiple sequence alignment,'' Proceedings of the National Academy of Sciences, Vol. 86, pp. 4412-4415, June 1989.

\bibitem[Monge, 1781]{monge1781} G. Monge, ``D\'{e}blai et remblai,'' M\'{e}moires de l'Acad\'{e}mie des Sciences, 1781.

\bibitem[Munkres, 1957]{munkres1957} J. Munkres, ``Algorithms for the Assignment and Transportation Problems,'' Journal of the Society of Industrial and Applied Mathematics, 5(1):32-38, 1957 March.

\bibitem[Nash, 1959]{nash1959} S. Nash, ``The Higher Derivative Test for Extrema and Points of Inflection,'' The American Mathematical Monthly, Vol. 66, No. 8, (Oct., 1959), pp. 709-713

\bibitem[Notredame, 2002]{notredame2002} C. Notredame, ``Recent progresses in multiple sequence alignment: a survey,'' Pharmacogenomics (2002) 3:131-144.

\bibitem[Papadimitriou, Steiglitz, 1998]{papadimitriousteiglitz1998}, \textit{Combinatorial Optimization: Algorithms and Complexity}, Dover Publications, Mineola, New York, 1998.

\bibitem[Rieke et al. 1997]{riekeetal1997} F. Rieke, D. Warland, R. de Ruyter van Steveninck, W. Bialek, \textit{Spikes: Exploring the Neural Code}, The MIT Press, Cambridge Massachusetts, 1997.

\bibitem[Roweis, Saul, 2000]{roweissaul2000} S. Roweis and L. Saul, ``Nonlinear dimensionality reduction by locally linear embedding,'' Science, New Series, Vol. 290, No. 5500, (Dec. 22, 2000): 2323-2326.

\bibitem[Samonds, Bonds, 2003]{samondsbonds2003} J. Samonds and A. Bonds, ``From another angle: differences in cortical coding between fine and coarse discrimination of orientation,'' Journal of Neurophysiology 91: 1193-1202, 2004.

\bibitem[Schrauwen, Campenhout, 2007]{schrauwencampenhout2007} B. Schrauwen, J. van Campenhout, ``Linking non-binned spike train kernels to several existing spike train metrics,'' Neurocomputing, Volume 70, Issues 7-9, March 2007, Pages 1247-1253.

\bibitem[Schrijver, 2003]{schrijver2003} A. Schrijver, \textit{A Course in Combinatorial Optimization}, available at http://www.math.ku.edu/$\sim$jmartin/old-courses/math996-F06/Schrijver.pdf

\bibitem[Sellers, 1974]{sellers1974} P. Sellers, ``On the theory and computation of evolutionary distances,'' SIAM Journal on Applied Mathematics, Vol. 26, No. 4 (Jun, 1974): 787-793.

\bibitem[Thompson, Higgins, Gibson, 1994]{thompsonhigginsgibson1994} J. Thompson, D. Higgins, and T. Gibson, ``CLUSTAL W: improving the sensitivity of progressive multiple sequence alignment through sequence weighting, position-specific gap penalties and weight matrix choice,'' Nucleic Acids Research, 1994, Vol. 22, No. 22, 4673-4680.

\bibitem[van Rossum, 2001]{vanrossum2001} M. van Rossum, ``A novel spike distance,'' Neural Computation 2001 Apr;13(4):751-63

\bibitem[Victor, 2005]{victor2005} J. Victor, ``Spike train metrics,'' Current Opinion in Neurobiology 2005, 15:585-592.

\bibitem[Victor, Purpura, 1996]{victorpurpura1996} J. Victor and K. Purpura, ``Nature and precision of temporal coding in visual cortex: a metric-space analysis,'' Journal of Neurophysiology, Vol. 76, No. 2, August 1996: 1310-1326.

\bibitem[Victor, Purpura, 1997]{victorpurpura1997} J. Victor and K. Purpura, ``Metric-space analysis of spike trains: theory, algorithms, and applications,'' Network 8, 127-164 (1997).

\end{thebibliography}
\end{document}